\documentclass{elsart1p}
\usepackage{epsfig}
\usepackage{graphicx}
\usepackage{amssymb}
\begin{document}
\begin{frontmatter}
%
%
%
%
%
\title{Production of antinuclei in pp collisions at $\sqrt{s}$ = 7 TeV with ALICE at the LHC}
%
%

\author{N. Sharma for the ALICE Collaboration}

\address{Department of Physics, Panjab University, Chandigarh, India}

\begin{abstract}
  First results of ALICE  on the production of nuclei and antinuclei in pp collisions at $\sqrt{s}$ = 7 TeV are presented. These particles are identified using the energy loss ($dE/dx$) measurements in the Time Projection Chamber. The Inner Tracking System allows a precise determination of the event vertex, by which primary and secondary particles are well separated. The high statistics of over 350 M events give a significant number of light nuclei and antinuclei such as (anti)deuterons, (anti)tritons, (anti)$^{3}\rm{He}$ and possibly (anti)hypertritons. The study of nuclei and antinuclei will help to understand their production mechanisms. Antinuclei production as a function of particle multiplicity in an event is discussed in this respect. Various particle ratios obtained from these collisions using predictions from a statistical model are also presented.
\end{abstract}

\end{frontmatter}

\section{Introduction}
It is believed that during the initial stage of the universe, matter and antimatter existed in equal abundance. It is still a mystery how this symmetry got lost in the evolution of the universe with no significant amount of antimatter being present. High energy collisions recreate energy density close to that of the universe microseconds after Big Bang. One of the striking features of particle production at these energies is the comparable abundance of matter and antimatter. Nuclei are abundant in the universe, but antinuclei heavier than antiproton have been observed only as products of interactions at particle accelerators~\cite{Alcaraz, Fuke}.  Recently, the STAR experiment at RHIC reported the observation of antimatter hypernuclei in AuAu collisions at $\sqrt{s_{NN}}$ = 200 GeV~\cite{Abelev:2010}.

 Production of antinuclei is possible via several mechanisms. Presumably the dominant mechanism for antinuclei production is via final-state coalescence. In this framework, the produced antinucleons merge to form light antinuclear clusters during freeze-out~\cite{Butler, Gutbrod}.

In this paper we present the first results on identified nuclei and antinuclei in mid-rapidity region for pp collisions at $\sqrt{s}$ = 7 TeV at the LHC. These particles are identified using the specific energy loss ($dE/dx$) measurements in the Time Projection Chamber (TPC) of the ALICE experiment.

\begin{figure}
\begin{center}
  \includegraphics[width=0.85\linewidth]{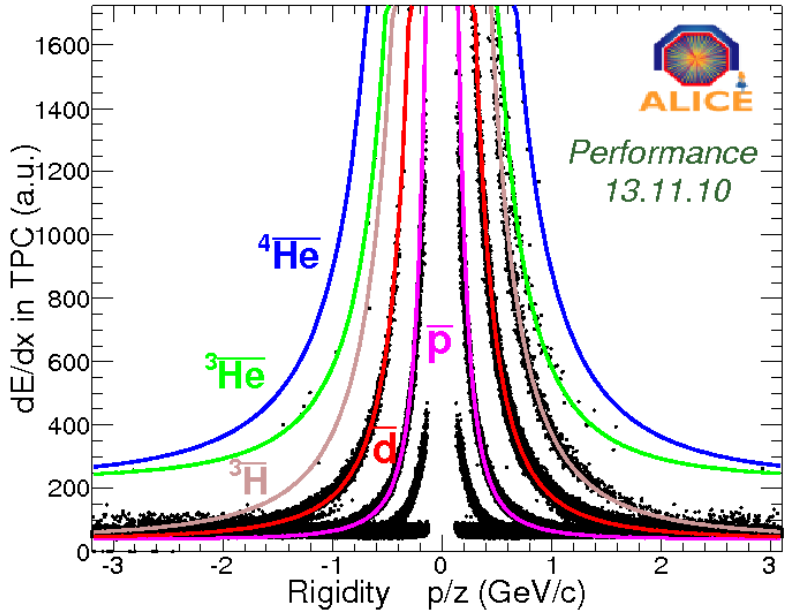}
\caption{Specific energy loss $dE/dx$ vs. momentum/charge of TPC tracks (colour online). The solid lines are parametrization of the Bethe-Bloch curve.}
\label{dedx}
\end{center}
\end{figure}

\section{Experiment}
\begin{figure}
\begin{center}
\includegraphics[scale=0.33]{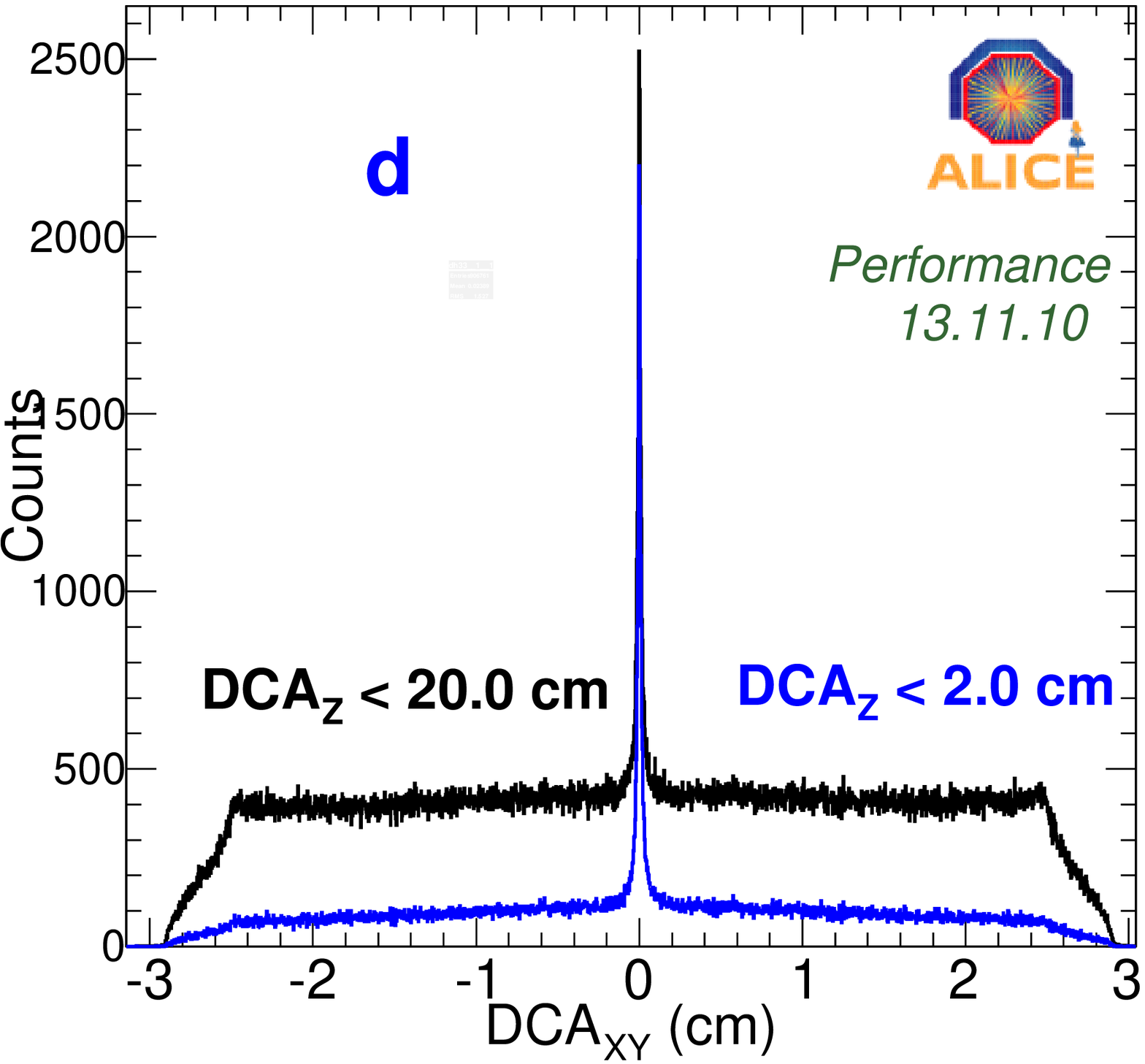}
\includegraphics[scale=0.31]{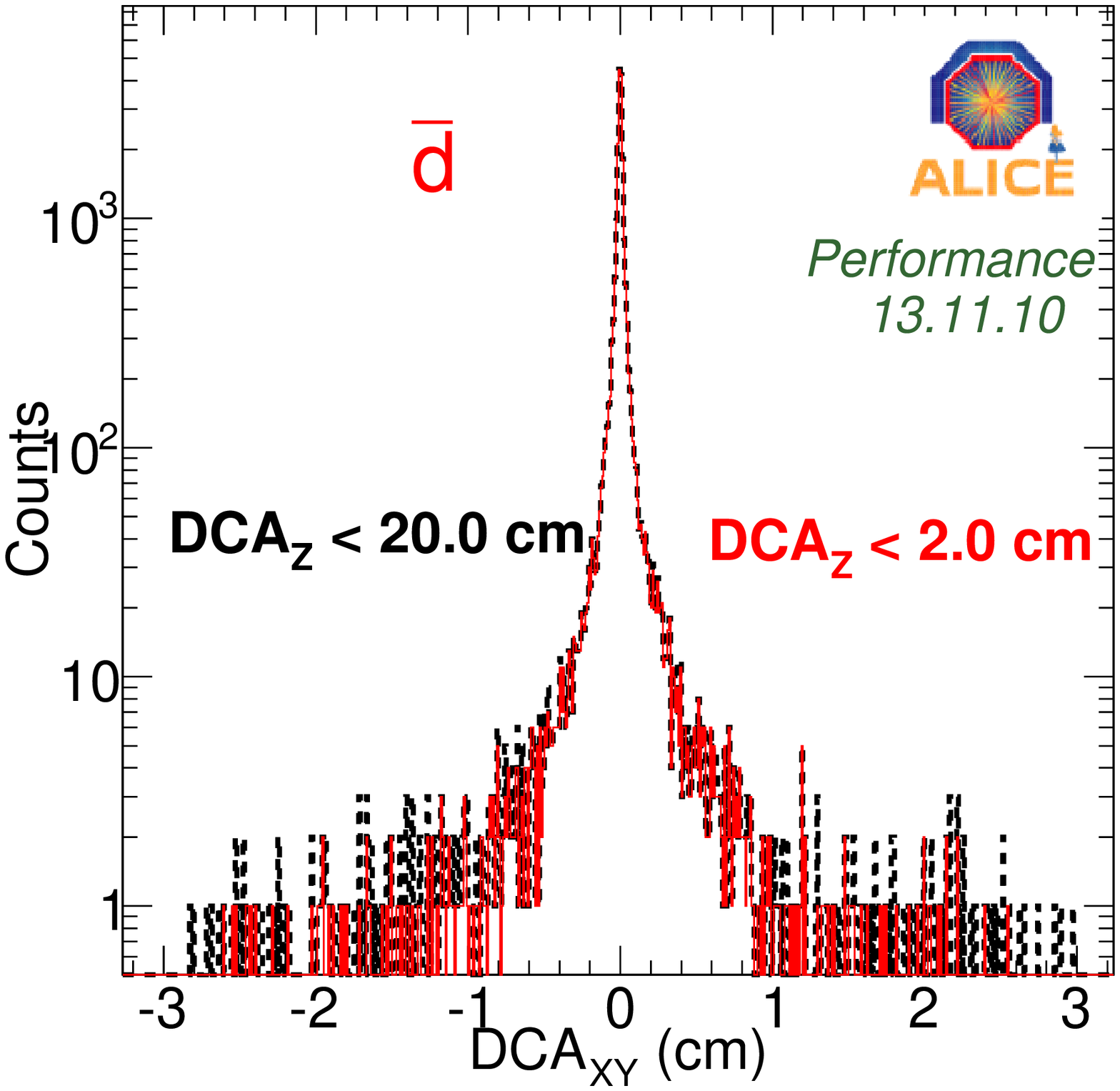}
\caption{$\rm{DCA}_{XY}$ distribution for deuterons and anti deuterons. Left panel: $\rm{DCA}_{Z}$ cut of 2.0 cm reduces large background for deuterons. 
Right panel: $\rm{DCA}_{Z}$ cut of 2.0 cm reduces background without affecting primary anti deuterons.}
\label{DCA_plot}
\end{center}
\end{figure}

The ALICE detector, its performance and description of its various subsystems are discussed in details in Ref. 
~\cite{Alessandro, Cinausero, Aamod, Aamodt:2010d}. Over $350$ M triggered events at $\sqrt{s}$ = 7 TeV are analyzed. For analysis we have used Time Projection Chamber (TPC) which has full acceptance of tracks with $|\eta|$ $<$ 0.9. The specific energy loss $dE/dx$ versus momentum/charge of TPC tracks are shown in Fig. ~\ref{dedx}. Nuclei and antinuclei like (anti)deuterons, (anti)tritons and (anti)$^{3}\rm{He}$ are clearly identified over wide range of momenta. The Inner Tracking System (ITS) containing six silicon layers, is used for precise determination of the event vertex, by which primary and secondary particles are well separated. Primary tracks are selected with the condition that, at least two clusters in the ITS are associated to the track. Secondary tracks are further rejected using the distance of closest approach (DCA) to the reconstructed primary vertex position information. The probability of antinuclei production by interaction of produced particles with material is very small. As can be seen from Fig.~\ref{DCA_plot} the $\rm{DCA}_{Z}$ (Z axis is along beam line) cut of 2.0 cm reduces a large fraction of background for deuterons ($\rm{d}$), but does not change the distribution for anti deuterons ($\rm\overline{d}$). Raw yield is obtained from the area in the peak minus background.

\begin{figure}
\begin{center}
\includegraphics[scale=0.31]{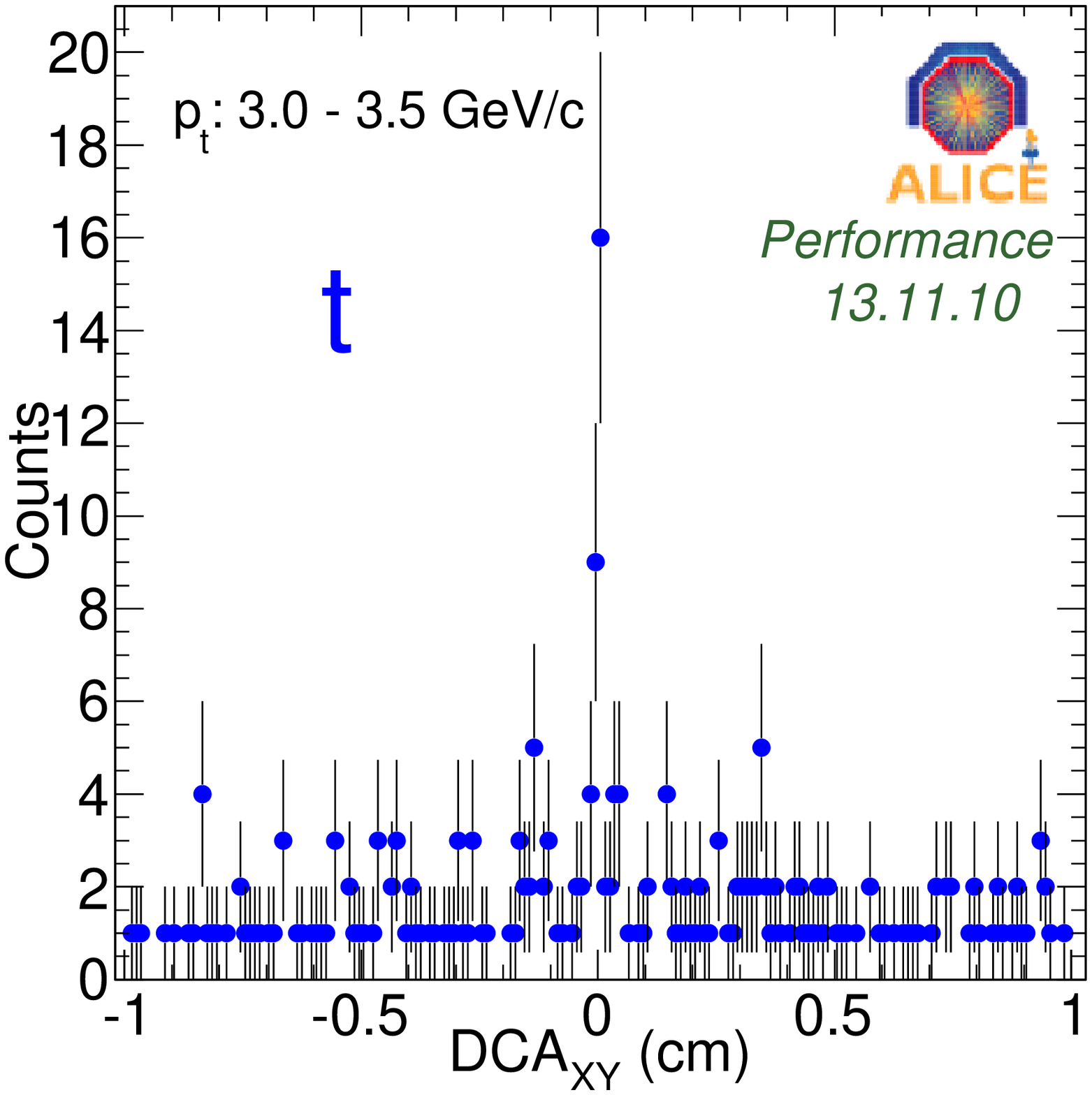}
\includegraphics[scale=0.31]{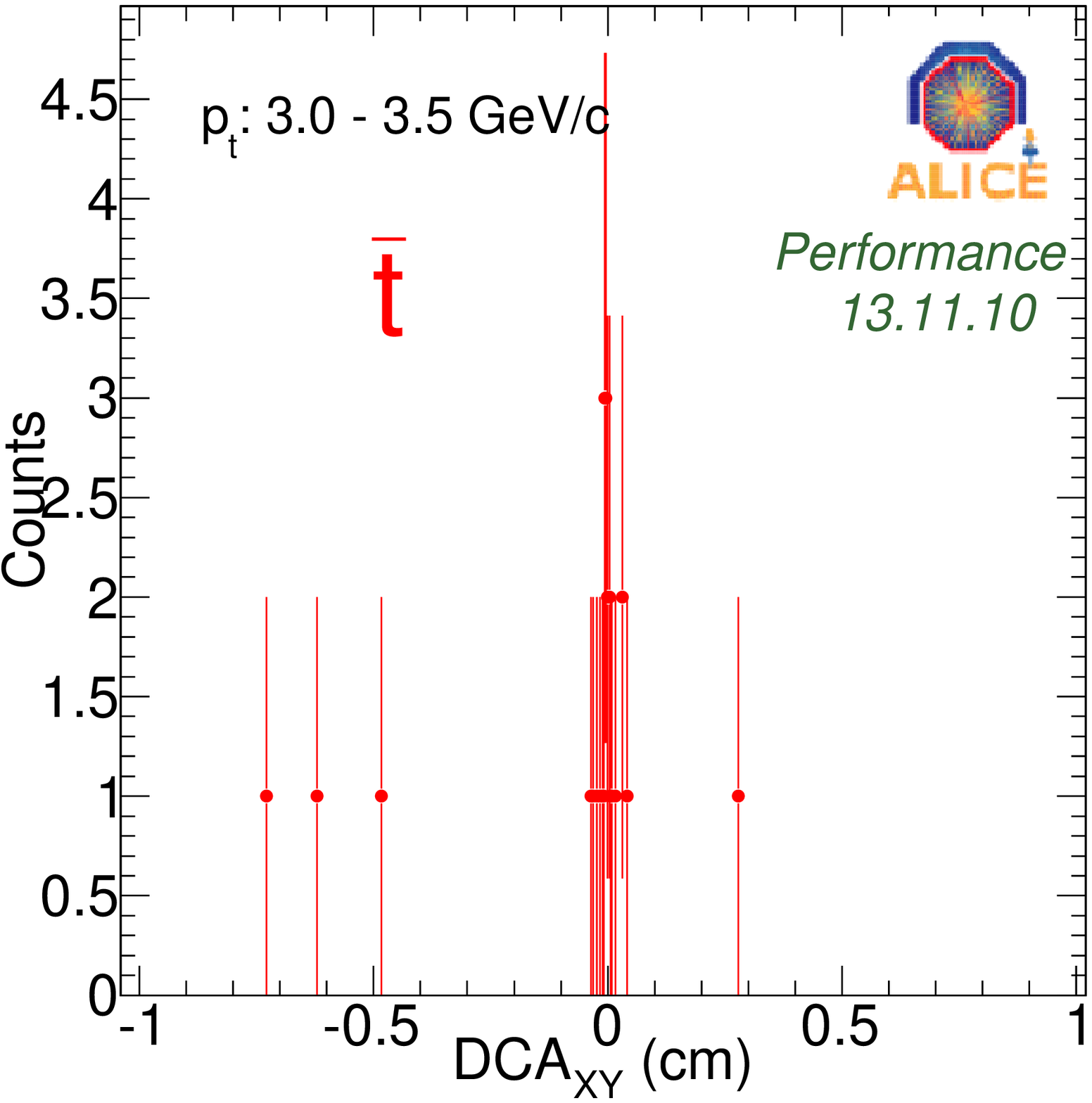}
\caption{$\rm{DCA}_{XY}$ distribution for tritons and anti tritons. $|\rm{DCA}_{XY}|$ $<$ 0.5 cm cut is chosen to get the raw yields of tritons and anti tritons in the transverse momentum range 3.0 GeV/$c$ $<$ $\ensuremath{p_{\rm t}}$ $<$ 3.5 GeV/$c$.}
\label{t_DCA_plot}
\end{center}
\end{figure}

Secondary tritons ($\rm{t}$) and anti tritons ($\rm\overline{t}$) are rejected with their $\rm{DCA}_{XY}$ information. Figure~\ref{t_DCA_plot} shows the $\rm{DCA}_{XY}$ resolution and a cut of $|\rm{DCA}_{XY}|$ $<$ 0.5 cm is chosen to get the raw yields of tritons and anti tritons. The same method is used to get the raw yields of Helium3 ($^{3}\rm{He}$) and anti Helium3 ($^{3}\rm\overline{He}$).

\section{Results}

\begin{figure}
\begin{center}
\includegraphics[scale=0.32]{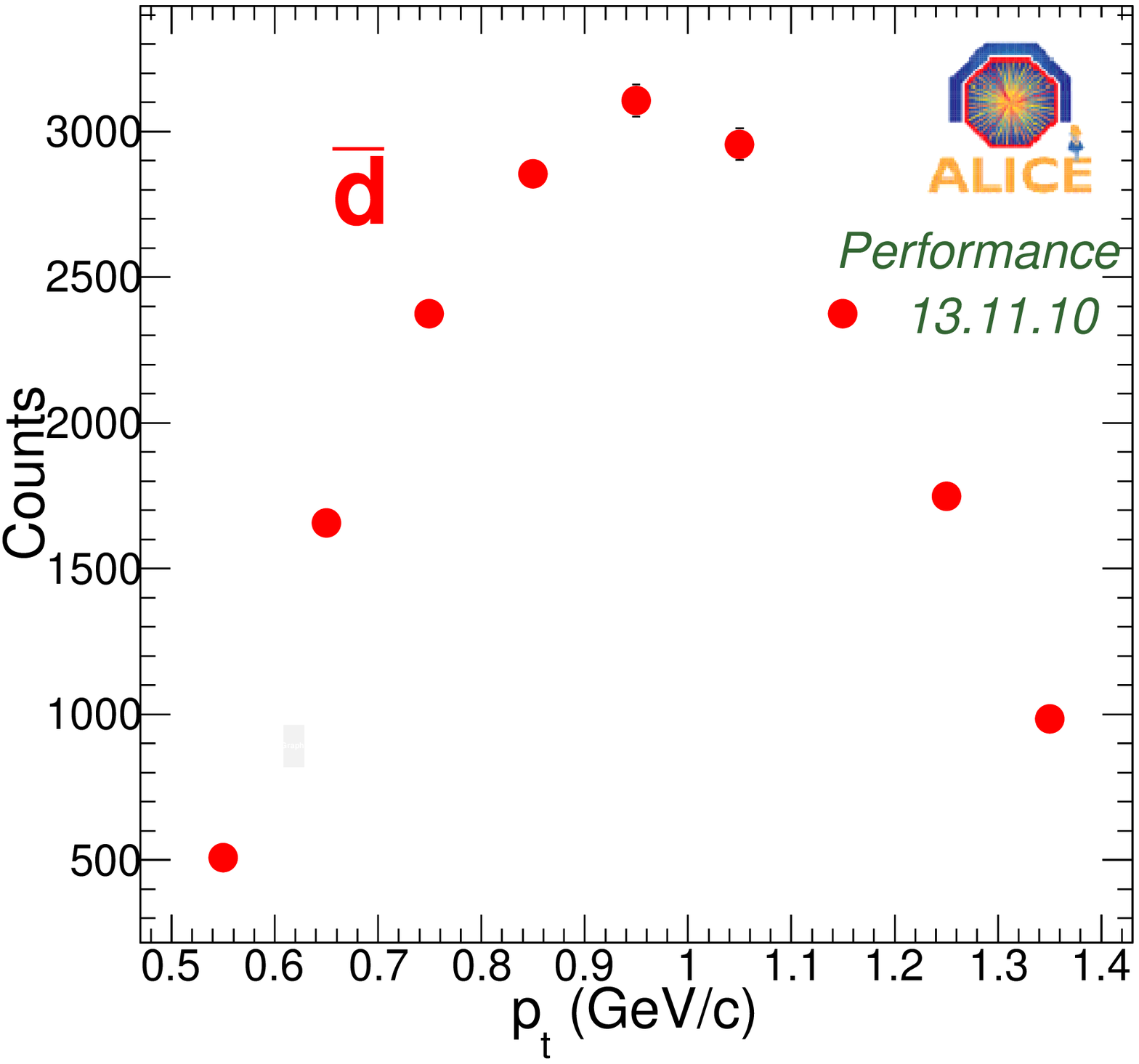}
\includegraphics[scale=0.34]{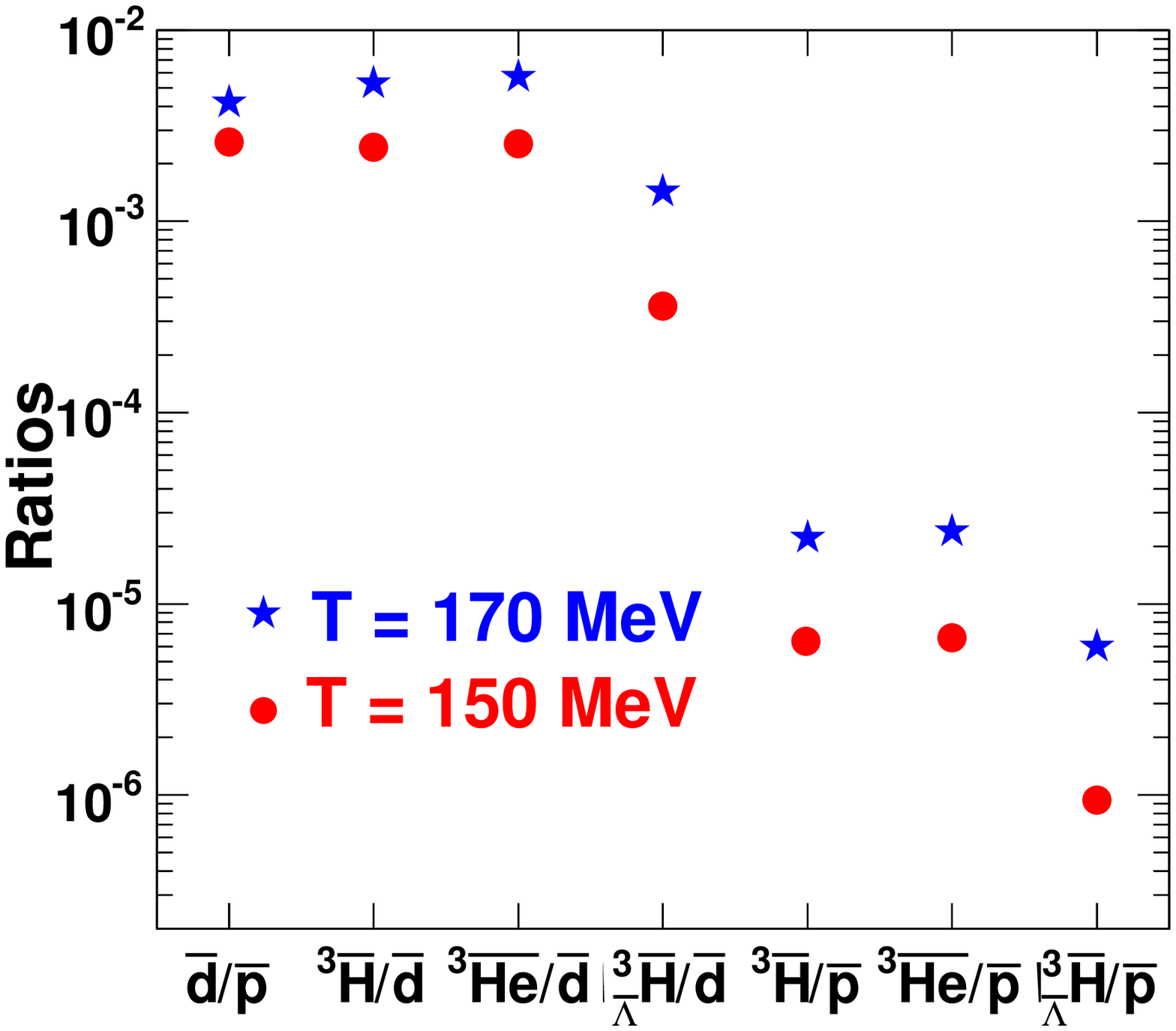}
\caption{Left Panel: Raw yield of anti deuterons as a function transverse momentum ($\ensuremath{p_{\rm t}}$)
Right Panel: Ratios of particles with different masses predicted using the Strangeness Canonical THERMUS model.}
\label{thermus}
\end{center}
\end{figure}




Deuterons and anti deuterons are identified in the transverse momentum range $0.5$ GeV/$c$ $< \ensuremath{p_{\rm t}} < 1.4$ GeV/$c$. The raw yield of anti deuterons is shown in the left panel of Fig.~\ref{thermus}. To get the final yields of nuclei and antinuclei the efficiency correction and annihilation correction have to be taken into account, this work is ongoing.


In order to estimate the yields of these complex nuclei we have used the statistical approach. Using the THERMUS code~\cite{Wheaton:2004qb}, calculations for pp collisions at $\sqrt{s}$ = 7 TeV have been carried out within the strangeness canonical approach. Ratio of particles with different masses are plotted in the right panel of Fig.~\ref{thermus} by assuming temperature ($T$) of 170 MeV and 150 MeV. This shows that the particle ratios are very sensitive to the freeze-out temperature.

The average track multiplicity for events with $\pi^{-}$, $K^{-}$, $\rm\overline{p}$, $\rm\overline{d}$, $\rm\overline{t}$ and $^{3}\rm\overline{He}$ are compared. It has been observed that massive particles have higher average track multiplicity. This could be taken as a hint for particle production via coalescence.

It will be interesting to study these nuclei and antinuclei in pp collisions at $\sqrt{s}$ = 7 TeV after efficiency and annihilation corrections. It will be more interesting to study the same in PbPb collisions at $\sqrt{s_{NN}}$ = 2.76 TeV using ALICE at the LHC.


\label{}

\end{document}